\documentclass[fdp,fleqn]{w-art}
\usepackage{times}
\usepackage{w-thm}
\usepackage[]{graphicx}
\usepackage[]{hyperref}
\usepackage{amsfonts}
\newcommand{\be}{\begin{equation}}
\newcommand{\ee}{\end{equation}}
\newcommand{\ba}{\begin{eqnarray}}
\newcommand{\ea}{\end{eqnarray}}

\newcommand{\e}{\epsilon}

\newcommand{\p}{\partial}

\newcommand{\nn}{\nonumber}

\newcommand{\lp}{\left(}
\newcommand{\rp}{\right)}

\begin{document}
\DOIsuffix{theDOIsuffix}
\Volume{55}
\Issue{1}
\Month{01}
\Year{2007}
\pagespan{1}{}

\hfill ITP-UH-04/10



\title[On Classical de Sitter Vacua in String Theory]{On Classical de Sitter Vacua in String Theory}


\author[T. Wrase]{Timm Wrase%
  }
\address[]{Institut f\"{u}r Theoretische Physik \& Center for Quantum Engineering and Spacetime Research\\
       Leibniz Universit\"{a}t Hannover, Appelstra{\ss}e 2, 30167 Hannover, Germany}
\author[M. Zagermann]{Marco Zagermann\footnote{Corresponding author. \quad E-mail:~\textsf{Timm.Wrase, Marco.Zagermann @itp.uni-hannover.de}}}
\begin{abstract}
We review the prospect of obtaining tree-level de Sitter (dS) vacua and slow-roll inflation models in string compactifications. Restricting ourselves to the closed string sector and assuming the absence of NSNS-sources, we classify the minimal classical ingredients that evade the simplest no-go theorems against dS vacua and inflation. Spaces with negative integrated curvature together with certain combinations of low-dimensional orientifold planes and low-rank RR-fluxes emerge as the most promising setups of this analysis. We focus on two well-controlled classes that lead to an effective 4D, $\mathcal{N}=1$ supergravity description: Type IIA theory on group or coset manifolds with SU(3)-structure and O6-planes, as well as type IIB compactifications on SU(2)-structure manifolds with O5- and O7-planes. While fully stabilized AdS vacua are generically possible, a number of problems encountered in the search for dS vacua are discussed.
\end{abstract}
\maketitle                   





\section{Introduction}
String compactifications with background fluxes (see e.g. \cite{Grana:2005jc,Douglas:2006es,Blumenhagen:2006ci,Denef:2007pq} for reviews) provide a simple framework in which the stabilization of moduli fields can be discussed in a very controlled and natural way. A complete stabilization of all moduli may also require the inclusion of quantum corrections, as e.g. in \cite{Kachru:2003aw}, but there are also scenarios where the fluxes alone are sufficient for a tree-level stabilization of all closed string moduli \cite{DeWolfe:2005uu}.

From a cosmological point of view, it is especially interesting to understand moduli stabilization at positive potential energy, either in order to obtain local dS minima so as to describe the present accelerated cosmic expansion, or in order to stabilize possible runaway directions in inflationary potentials. A particularly well controlled situation would be one in which this could be achieved at a purely classical level, i.e., by the dimensional reduction of the standard two-derivative 10D supergravity action supplemented with the lowest order actions for brane type sources.

As is well-known for a long time, however, there are powerful no-go theorems \cite{Gibbons:1984kp,deWit:1986xg,Maldacena:2000mw} that forbid such tree-level dS compactifications under a few simple assumptions, one of them being the absence of negative tension objects such as orientifold planes. As orientifold planes are a common ingredient in phenomenologically interesting type II compactifications, it seems natural to explore the possibility of tree-level dS vacua or inflation models in type II orientifolds. It is the purpose of this note to give an overview of the most promising controlled models of this type. For simplicity, we do not consider D-branes and the associated open string moduli (although the analysis would be similar). Moreover, we take the O-planes to be smeared over their transverse directions \cite{Grimm:2004ua,DeWolfe:2005uu,Koerber:2008rx,Caviezel:2008ik}, assuming that the results approximate fully localized warped solutions \cite{Acharya:2006ne} consistent with the results of \cite{Douglas:2010rt}.

\section{No-go theorems in the volume-dilaton plane}
Constructions of dS vacua or inflation models from classical string compactifications are severely limited by a set of very simple ``no-go theorems''. These no-go theorems go beyond \cite{Gibbons:1984kp,deWit:1986xg,Maldacena:2000mw}, as they do allow for orientifold planes and generalize the theorems used in \cite{Hertzberg:2007wc} for IIA flux compactifications on Calabi-Yau spaces with O6-planes and the IIB setup of \cite{Giddings:2001yu}. They follow from the scaling behavior of the different scalar potential contributions with respect to two universal moduli fields that occur in any perturbative and geometric string compactification. These two fields are the volume modulus $\rho \equiv (\textrm{vol}_6)^{1/3}$ and an orthogonal modulus related to the dilaton: $\tau \equiv e^{-\phi} \sqrt{\textrm{vol}_6} = e^{-\phi} \rho^{3/2}$, where $\phi$ denotes the 10D dilaton, and $\textrm{vol}_6$ is the 6D volume in string frame. After going to the four-dimensional Einstein frame, one then finds the following scalings for the contributions to the 4D scalar potential coming from the $H$-flux, the RR-fluxes $F_p$, as well as from O$q$-planes and the six-dimensional Ricci scalar $R_6$:
\be
V_H \sim \tau^{-2} \rho^{-3}, \quad V_{F_p} \sim \tau^{-4} \rho^{3-p}, \quad V_{Oq} \sim \tau^{-3} \rho^{\frac{q-6}{2}}, \quad V_{R_6} \sim \tau^{-2} \rho^{-1}. \label{Scalings}
\ee
Note that $V_H, V_{F_p} \geq 0$ and $V_{Oq} \leq 0$ while $V_{R_6}$ can have either sign.

The most widely studied classes of compactifications are based on special holonomy manifolds such as $CY_3$'s, $T^2 \times K3$ or $T^6$, which are Ricci-flat, i.e. they have $V_{R_6}=0$. In order to find the minimal necessary ingredients for classical dS vacua in this simplified case, we act on $V=V_H + \sum_p V_{F_p} + \sum_q V_{Oq}$ \footnote{The possible values for $p$ and $q$ consistent with four-dimensional Lorentz invariants are $p\in\{0,2,4,6\}$, $q\in\{4,6,8\}$ in type IIA theory and $p\in\{1,3,5\}$, $q\in \{3,5,7,9\}$ in type IIB theory. To cancel the charges of the O$q$-planes in the Ricci-flat case we need $V_H \neq 0$ and $\sum_p V_{F_p}\neq 0$. For compactification with $V_{R_6} \neq 0$ we need $\sum_p V_{F_p}\neq 0$.} with a differential operator $D:= -a \tau \p_\tau - b \rho \p_\rho$, where $a$ and $b$ denote some as yet arbitrary real constants. If one can show that there is a constant $c>0$ such that $D V \geq c V$, then dS vacua and generically slow-roll inflation are excluded. Indeed, a dS extremum requires $D V=0$ and $V>0$, which is inconsistent with $D V \geq c V >0$. Similarly, the slow-roll parameter $\e =\frac{K^{i\bar{j}} \p_i V \p_{\bar{j}} V}{V^2} \geq \frac{c^2}{4a^2+3b^2}$ is normally of order one so that slow-roll inflation with $\e \ll 1$ is forbidden. Using (\ref{Scalings}), this means that, if we can find $a,b$ such that
\ba
&&D V_H = (2a+3b) V_H, \quad D V_{F_p} = (4a+(p-3)b) V_{F_p}, \quad D V_{Oq} = \lp 3a + \frac{6-q}{2} b\rp V_{Oq}, \nn\\
&& \text{with} \quad (2a+3b) \geq c \geq \lp 3a + \frac{6-q}{2} b\rp, \quad (4a+(p-3)b) \geq c \geq \lp 3a + \frac{6-q}{2} b\rp, \quad \forall p,q, \nn
\ea
then we have a no-go theorem that forbids classical dS vacua and inflation. The two inequalities above have a solution if and only if $q+p-6\geq 0,\, \forall p,q$. This condition is for example satisfied for type IIA compactifications on a $CY_3$ with O6-planes and arbitrary RR-fluxes or, analogously, for the type IIB theory with O3-planes and $F_3$-flux \cite{Hertzberg:2007wc}. Conversely, avoiding this no-go theorem at the classical level would require compactifications with $H$-flux that, in type IIA, allow for O4-planes and $F_0$-flux or, in type IIB, allow for O3-planes and $F_1$-flux. However, the $F_0$-flux needs to be odd under the O4 orientifold projection and therefore normally has to vanish. Similarly, all one-forms are normally odd under the O3 orientifold projection, but the $F_1$-flux has to be even and should therefore also vanish in this constellation\footnote{It might in principle be possible to consider compactifications on toroidal orbifolds that have for example $F_1$-flux only in the bulk and O3-planes in the twisted sector. In this note we restrict ourselves to the bulk sector only.}.

A possible way out of these difficulties might be to allow also for non Ricci-flat manifolds. This would contribute the additional term $V_{R_6} \sim -R_6 \sim \tau^{-2} \rho^{-1}$ to the scalar potential. It is easy to check that for positively curved manifolds ($V_{R_6} < 0$) the above conditions cannot be relaxed. Although $H$-flux is not necessary anymore to cancel the O$q$-plane tadpole, one still needs it to avoid a no-go theorem with $b=0$. For manifolds with integrated negative curvature, on the other hand, the condition for a no-go theorem becomes relaxed to $q+p-8\geq 0,\, \forall p,q$. The only exception is the case with O3-planes and $F_5$-flux, which saturates this inequality, but would require $c=0$ and therefore cannot be excluded based on this inequality. Table \ref{table} summarizes the no-go theorems against classical dS vacua and slow-roll inflation\footnote{In \cite{Danielsson:2009ff} a similar conclusion is reached for dS vacua with small cosmological constant using the ``abc''-method of \cite{Silverstein:2007ac}.}.

\begin{table}[h!]
\begin{center}
\begin{tabular}{|c|c|c|c|}
  \hline
  Curvature & No-go, if & No no-go in IIA with & No no-go in IIB with \\ \hline \hline
  $V_{R_6} \sim -R_6 \leq 0$ & \begin{tabular}{c} $q+p-6\geq 0,\, \forall p,q,$\\ $\e \geq \frac{(3+q)^2}{3+q^2} \geq \frac{12}{7}$ \end{tabular} & O4-planes and $H$, $F_0$-flux & O3-planes and $H$, $F_1$-flux  \\ \hline
  $V_{R_6} \sim -R_6 > 0$ & \begin{tabular}{c}
                            $q+p-8\geq 0,\, \forall p,q,$ \\
                            (except $q=3,p=5$)\\ $\e \geq \frac{(q-3)^2}{q^2-8q+19} \geq \frac{1}{3}$ \end{tabular} & \begin{tabular}{c}
                                                             O4-planes and $F_0$-flux \\
                                                             O4-planes and $F_2$-flux \\
                                                             O6-planes and $F_0$-flux
                                                           \end{tabular} &
                                                           \begin{tabular}{c}
                                                             O3-planes and $F_1$-flux \\
                                                             O3-planes and $F_3$-flux \\
                                                             O3-planes and $F_5$-flux \\
                                                             O5-planes and $F_1$-flux
                                                           \end{tabular} \\ \hline
\end{tabular}
\end{center}
\caption{The table summarizes the conditions that are needed in order to find a no-go theorem in the $(\rho,\tau)$-plane and the resulting lower bound on the slow-roll parameter $\e$. The third and fourth column spell out the minimal ingredients necessary to evade such a no-go theorem.\label{table}}
\end{table}

As we have argued above, it is difficult to find explicit examples with O3-planes and $F_1$-flux or with O4-planes and $F_0$-flux. The same turns out to be true for O3-planes with non-vanishing curvature \cite{Caviezel:2009tu}. The prospects of stabilizing all moduli at tree-level in IIA compactifications with O4-planes are not clear so we will restrict ourselves in the rest of these notes to compactifications on manifolds with negative curvature and O6-planes in type IIA or O5-planes in type IIB. Moreover, we will focus on those compactifications that give an effective 4D,  $\mathcal{N}=1$ supergravity action, which leads us to SU(3)-structure manifolds with O6-planes in IIA, and  SU(2)-structure compactifications with O5- and O7-planes in type IIB string theory.\footnote{We need to compactify on an SU(2)-structure manifold in IIB, because the $F_1$-flux requires a 1-form. $\mathcal{N}=1$ supersymmetry then also requires O7-planes in addition to the O5-planes.}

\section{Type IIA}
The attempts to construct classical dS vacua in IIA compactifications on manifolds with negative curvature and O6-planes were initiated in \cite{Silverstein:2007ac}, where also other types of sources such as KK5-monopoles were used. A similar construction with only the ingredients of eq. \eqref{Scalings} was attempted in \cite{Haque:2008jz}, whereas in \cite{Danielsson:2009ff} the authors argued that the constructions of \cite{Silverstein:2007ac} and \cite{Haque:2008jz} cannot be lifted to full 10D solutions.

In this note, we review IIA compactifications on a special class of SU(3)-structure manifolds, namely coset spaces \cite{Koerber:2008rx,Caviezel:2008ik,Caviezel:2008tf} involving semisimple and Abelian groups, as well as twisted tori (solvmanifolds) \cite{Ihl:2007ah,Flauger:2008ad}. The underlying Lie group structure endows these spaces with a natural expansion basis (the left-invariant forms)
for the various higher-dimensional fields and fluxes, and one expects that the resulting 4D, $\mathcal{N}=1$ theory is a consistent truncation of the full 10D theory \cite{Cassani:2009ck}. Furthermore, in these compactifications it is possible to stabilize all moduli in AdS vacua \cite{Grimm:2004ua,DeWolfe:2005uu,Ihl:2007ah}. This means that the scalar potential generically depends on all moduli, which is a prerequisite for the construction of metastable dS vacua.

Whereas the previous analysis focused on the behavior of the potential in the volume-dilaton plane, it is clear that once the no-go theorems using these fields are circumvented, one must still make sure that there are no other steep directions of the scalar potential in directions outside the $(\rho,\tau)$-plane. For the coset spaces and twisted tori studied in \cite{Caviezel:2008tf,Flauger:2008ad}, the volume turns out to factorize further into a two-dimensional and a four-dimensional part: $\textrm{vol}_6 = \rho^3 = \rho_{(2)} \rho^2_{(4)}$. In such cases one can then study directions involving $\rho_{(2)}$ or $\rho_{(4)}$ and finds that, if for a given model
\be
(-2 \tau\p_\tau -\rho_{(4)} \p_{\rho_{(4)}}) V_{R_6} \geq 6 V_{R_6},
\ee
then the full scalar potential also satisfies $(-2 \tau\p_\tau -\rho_{(4)} \p_{\rho_{(4)}}) V \geq 6 V$, and one obtains the bound $\e \geq2$. In \cite{Caviezel:2008tf} six out of seven coset spaces could be excluded by this refined no-go theorem. In \cite{Flauger:2008ad} many similar no-go theorems were discussed and used to exclude almost all concrete models of twisted tori.

The only spaces that could not be excluded in this manner are $SU(2) \times SU(2)$ with four O6-planes and a twisted version of $T^6/\mathbb{Z}_2 \times \mathbb{Z}_2$. These two spaces are closely related \cite{Aldazabal:2007sn}, and therefore it is not surprising that they have very similar properties. In particular, for  both of these models it is possible to find (numerical) dS extrema \cite{Caviezel:2008tf,Flauger:2008ad}. Unfortunately, these dS extrema are unstable as one of the 14 real scalar fields turns out to be tachyonic with an $\eta$ parameter of order one. Interestingly, this tachyon is not the potential tachyon candidate identified for certain types of K\"ahler potentials in \cite{Covi:2008ea}. This can also be seen from the results in \cite{deCarlos:2009fq}, where a similar K\"ahler potential and a modified superpotential based on non-geometric fluxes lead to stable dS vacua (see also \cite{Dall'Agata:2009gv,Roest:2009tt,Dibitetto:2010rg}).

\section{Type IIB}
For type IIB compactifications we have seen that it is possible to evade simple no-go theorems in the $(\rho,\tau)$-plane if one includes O5-planes and $F_1$-flux. A concrete class of compactifications that allows for these ingredients and also preserves $\mathcal{N}=1$ supersymmetry in 4D was presented in \cite{Caviezel:2009tu} based on 6D SU(2)-structure spaces with O5- and O7-planes. As discussed there, these compactifications are formally T-dual to compactification of type IIA on SU(3)-structure spaces with O6-planes, however these T-duals are generically non-geometric and hence do not fall under the analysis of the previous section.

This particular class of IIB compactifications has the very interesting property that the tree-level scalar potential allows for fully stabilized supersymmetric AdS vacua with large volume and small string coupling \cite{Caviezel:2009tu}. This is very different from the no-scale property of classical type IIB compactifications on $CY_3$ manifolds along the lines of \cite{Giddings:2001yu}. It also shows that the scalar potential generically depends on all moduli.

For six-dimensional SU(2)-structure spaces the split of the volume $\textrm{vol}_6 = \rho^3 = \rho_{(2)} \rho^2_{(4)}$ into a two-dimensional and a four-dimensional part is very natural, and also the complex structure moduli naturally split into two classes. This allows one \cite{Caviezel:2009tu} to derive many no-go theorems and exclude most concrete examples of coset spaces and twisted tori with SU(2)-structure. The only space that was found to evade all the no-go theorems is $SU(2) \times SU(2)$ with an SU(2)-structure and O5- and O7-planes. Just as in the IIA analogue, we can find dS critical points, but again these have at least one tachyonic direction with a large $\eta$ parameter. It would be interesting to understand the geometrical meaning of this tachyon as well as the relation of the dS extrema found in \cite{Caviezel:2008tf,Flauger:2008ad,Caviezel:2009tu} to fully localized warped 10D solutions \cite{Acharya:2006ne,Douglas:2010rt}.

\begin{acknowledgement}
This work was supported by the German Research Foundation (DFG) within the Emmy Noether Program (Grant number ZA 279/1-2) and the Cluster of Excellence ``QUEST".
\end{acknowledgement}

\end{document}